\documentclass{Interspeech2024}

\interspeechcameraready 

\usepackage{amsmath,graphicx,multirow}
\usepackage{algorithm}
\usepackage{algpseudocode}
\usepackage{graphicx} 
\usepackage{caption} 
\usepackage{subcaption} 
\usepackage{xcolor}
\usepackage{amssymb}
\usepackage{dsfont}
\usepackage{comment}
\usepackage{graphicx}
\usepackage{multicol}
\usepackage{caption}
\begin{document}
\title{Target Speaker Extraction with Curriculum Learning}
\name[affiliation={1,2}]{Yun}{Liu}
\name[affiliation={1}]{Xuechen}{Liu}
\name[affiliation={3}]{Xiaoxiao}{Miao}
\name[affiliation={1,2}]{Junichi}{Yamagishi}
\address{
  $^1$National Institute of Informatics, Tokyo, Japan\\
  $^2$Sokendai, Kanagawa, Japan
  $^3$Singapore Institute of Technology, Singapore }
\email{{yunliu,xuecliu,jyamagis}@nii.ac.jp}
\maketitle
\begin{abstract}
This paper presents a novel approach to target speaker extraction (TSE) using Curriculum Learning (CL) techniques, addressing the challenge of distinguishing a target speaker's voice from a mixture containing interfering speakers. For efficient training, we propose designing a curriculum that selects subsets of increasing complexity, such as increasing similarity between target and interfering speakers, and that selects training data strategically. Our CL strategies include both variants using predefined difficulty measures (e.g.\ gender, speaker similarity, and signal-to-distortion ratio) and ones using the TSE's standard objective function, each designed to expose the model gradually to more challenging scenarios. Comprehensive testing on the Libri2talker dataset demonstrated that our CL strategies for TSE improved the performance, and the results markedly exceeded baseline models without CL about 1 dB. 
\end{abstract}
\section{Introduction}
\label{sec:intro}
Target speaker extraction (TSE) technology, which isolates the speech of a specific speaker from a mixture of interfering speakers and noise, is essential in speech separation and is currently making substantial progress for improved telecommunication, hearing aids, and automatic speech recognition (ASR) systems. Despite progress with deep neural network (DNN) models, TSE is still known to be challenging, especially when speakers have similar characteristics. Even the latest end-to-end training and advanced network architectures often struggle with real-world audio complexities \cite{Zmolikova_Spkbeam_STSP19,Wang2019}. Thus several attempts have been made through, e.g., \ metric learning \cite{fu2019metricgan} and post-filtering strategies \cite{zhao2022target}. Our study remedies these shortcomings in TSE by incorporating curriculum learning (CL) \cite{wang2021survey,soviany2022curriculum} into its model training process.

CL imitates human learning by training machine learning models on tasks that gradually increase in complexity. This approach has been effective in computer vision (CV) \cite{Guo_2018_ECCV,jiang2014easy} and natural language processing \cite{platanios-etal-2019-competence,tay-etal-2019-simple}. CL is also closely related to progressive learning \cite{Progressivenina}. In a narrow sense, CL focuses on actively selecting training subsets with a suitable difficulty measure of training samples. In contrast, progressive learning focuses on continuous learning and gradually increasing the model capacity to handle the challenging conditions, although the two terms are sometimes used without distinction. EfficientNetV2 \cite{tan2021efficientnetv2}, a well-known architecture in CV, uses a progressive training procedure that gradually increases the size of the input images during training. Progressive learning has been investigated in the past for speech enhancement \cite{li2020speech,gao16_interspeech}. \cite{Ranjan2018ASRCL} proposed signal-to-noise ratio (SNR) values of training utterances in noise as a difficulty measure to select training data for ASR. High SNR values, which signify clearer and easier examples, have been hypothesized to  be more informative for estimating latent variables in a supervised learning context. 

For efficient training of TSE models that can handle challenging conditions, we propose designing a curriculum that selects subsets of increasing complexity, such as increasing similarity between target and interfering speakers, and performing progressive learning, in which a seed model trained on a subset containing “easy” samples is continuously trained to improve performance in TSE. Our CL strategies for the subset selection include both variants using predefined difficulty measures (gender, speaker similarity, signal-to-distortion ratio) and ones using the TSE's standard objective function, each designed to expose the model gradually to more challenging data. 

This paper is structured as follows: Section 2 reviews the task and architecture used for our TSE CL learning. Multiple proposed CL strategies are described in Section 3. Section 4 shows the experimental conditions and results, and Section 5 concludes our findings. 

\section{Task and Architecture}
\subsection{Task Definition}
\label{sec:format}

First, we briefly explain the objective of the TSE task. 
Given a reference waveform $\boldsymbol{w}^{(r_t)} = (w_{1_t}, \cdots, w_{N_{r_t}})$ containing speech signals of target speaker $t$ with $N_{r_t}$ sampling points, which is used to extract a speaker embedding $\boldsymbol{e}_t = E(\boldsymbol{w}^{(r_t)})$ via a neural speaker encoder $E$, TSE aims to output the estimated clean speech signals $\hat{w}^{(t)} = (\hat{w}_{1}, \cdots, \hat{w}_N)$ of the target speaker only from a given $N$ sampling points mixture $m = (m_1, \cdots, m_N) = (s_1 + s'_1 , \cdots, s_N + s'_N) $ that contains target speaker $t$'s clean speech sample $s_n$, interference speakers' signal $s'_n$ That is, 
\begin{equation}
\hat{w}^{(t)} = \textrm{TSE}(m, \boldsymbol{e}_t = E(\boldsymbol{w}^{(r_t)}) ; \theta)
\end{equation}
where $\theta$ is the model parameters.

\begin{figure}[!htb]
    \centering
    \includegraphics[width=1\columnwidth]{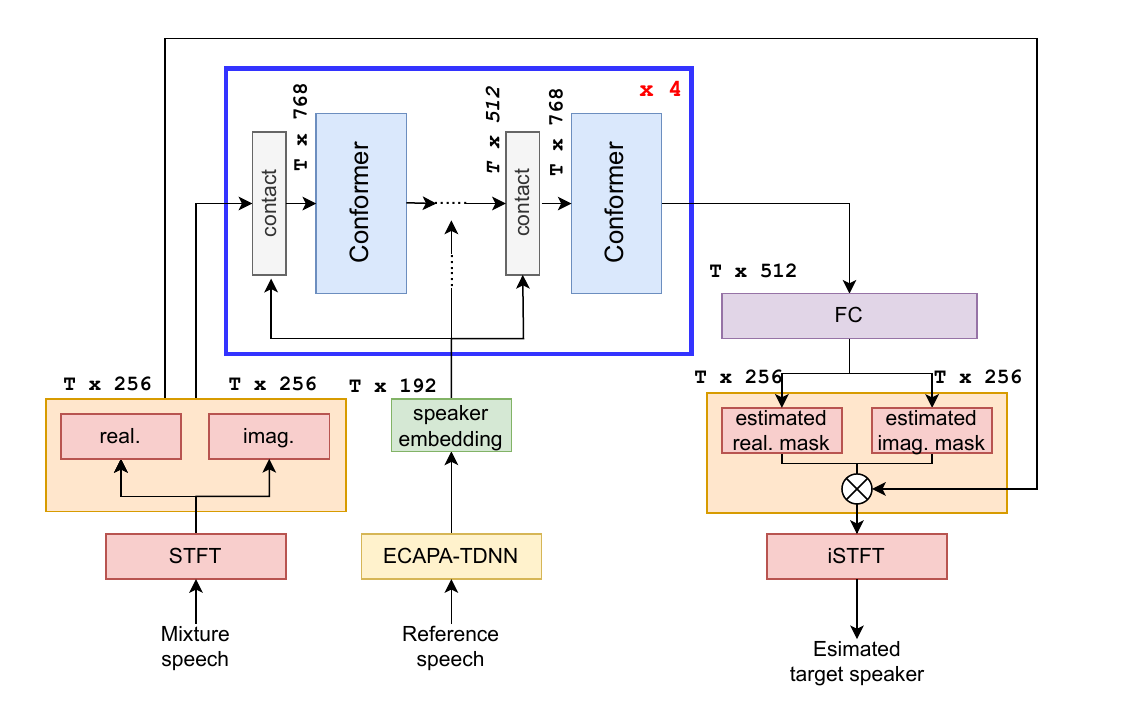}
    \caption{The conformer-based TSE architecture. Network details and loss function can be found in Supplementary materials.}
    \label{fig:architecture}
    \vspace{-0.6cm}
\end{figure}

\subsection{TSE Model Architecture}

The theme of this paper is CL through data selection and progressive learning of the TSE model $\theta$. Therefore, it can be adapted to any DNN architecture. Here we describe the structure of a modern TSE using Conformer \cite{sinha2022speaker} and the ratio masking of the complex-valued spectrum \cite{cirm,hu20g_interspeech} as an example. The TSE architecture is shown in Fig. \ref{fig:architecture}. The input sequence is a time-domain waveform of the mixture $m$, and the real and imaginary parts of the short-term Fourier transform (STFT) spectral sequence are fed into the main conformer blocks. 

The main body part has stacked conformer blocks to extract both local and global context features while reducing overall parameters~\cite{gulati2020conformer}. Each conformer block comprises two feed-forward blocks, a multi-head self-attention block, and a convolutional block. The initial input of the conformer block combines these STFT components with the speaker embedding $\boldsymbol{e}_t$ extracted from the reference waveform $\boldsymbol{w}^{(r_t)}$ of the target speaker $t$ using a neural speaker encoder $E(\boldsymbol{w}^{(r_t)})$, such as ECAPA-TDNN \cite{desplanques2020ecapa}. Meanwhile, the inputs to the subsequent conformer blocks merge the output of the previous block with the speaker embedding. Then, an additional fully-connected (FC) layer follows each conformer block, ensuring output dimensions equal to the sum of the real and imaginary parts of the STFT spectra. The estimated real and imaginary masks are obtained by splitting the output of the final FC layer. The final prediction is the real and imaginary parts of complex masking ratios \cite{cirm} to be multiplied by the complex spectrum of the input mixture $m$. Finally, it yields the waveform of the target speaker $t$'s clean speech $\hat{w}^{(t)}$ via the inverse STFT operation.

\section{Curriculum Learning Strategy for Target Speaker Extraction}
\label{sec:pagestyle}

As noted, the CL for TSE that we  investigated is a learning strategy in which the model is trained by progressively changing the training data from a subset containing more data that are easy to solve to a subset containing more complex data for which finding the correct solution is more difficult. In designing such a curriculum, two important aspects need to be considered: 1) Identifying training samples are inherently more difficult than other samples; 2) Determining the optimal timing and extent to which more complex data should be introduced. The first point requires a difficulty measure that represents the relative difficulty of extracting the target speaker from each training data sample; the second point requires an appropriate training scheduler that adjusts the order and timing of introducing more complex data subsets during model training. 
\vspace{-1mm}
\subsection{Difficulty measures}

\subsubsection{Gender}

Pilot experiments showed that the initial TSE model tended to be error-prone when the target and interfering speakers were of the same gender. Therefore, the gender information of the target and interrupting speakers could also be used as a difficulty measure. Possibilities are using speaker pairs in which the target speaker and the interfering speaker are of different genders in the early stages of learning, and  introducing additional speaker pairs in which the target speaker and the interfering speaker are of the same gender in the latter stages of learning.

\subsubsection{Speaker similarity} 

Another finding from pilot experiments was that the initial TSE model is error-prone when the target and interfering speakers are similar. We observed that when the cosine distance of the target speaker's and  the interfering speaker's embedding vectors is closer, the SNR improvement using the initial TSE model is relatively smaller. Therefore, the speaker similarity between the target speaker and the interfering speaker $\rm{sim}= \cos \left( E \left(s_1, \cdots, s_N \right), E \left(s'_1, \cdots, s'_N \right) \right) $ may also be used as a difficulty measure. A possible scheduling strategy is to use speaker pairs in which the target and interfering speakers are dissimilar to each other in accordance with their speaker embedding vectors in the early stages of learning and to use speaker pairs in which target and interfering speakers have more similar speaker characteristics in the latter stages of learning.
\vspace{-1mm}
\subsubsection{Signal-to-distortion ratio (SDR)} 

Because TSE training often uses an artificial mixture of the target speaker's speech and the interfering speaker's speech, the average of the SDR of the target speaker's signal to the interference signals can often be directly calculated, that is, $\rm{SDR} = \frac{1}{N} \sum_{n=1}^N 10\log_{10}\left(\frac{\lVert s_n \rVert^2}{\lVert s'_n  \rVert^2}\right)$. This can be used as a difficulty measure for each training sample. Lower SDRs indicate that the target speaker would be more difficult to extract due to distortion caused by the interfering speaker and noise.

\vspace{-1mm}
\subsubsection{SNR} 

Because our proposed method trains the model gradually, the initial seed model can be used to extract the target speaker's signal, and the resulting SNRs may also be used as a difficulty measure, that is, $\rm{SNR} = \frac{1}{N} \sum_{n=1}^N 10\log_{10}\left(\frac{\lVert s_n \rVert^2}{\lVert \hat{w}_n - s_n \rVert^2}\right)$. Lower SNRs indicate that the target speaker would be more difficult to extract because the initial seed model can not handle them properly. 
\vspace{-1mm}
\subsubsection{Self-paced} 

The aforementioned  four measures are designed by empirical knowledge or based on models learned in the past. We also propose a more dynamic measure based on a model that is being trained. This measure, called ``self-paced'' uses the training loss of the current TSE model, such as SNR and SI-SDR losses~\cite{convtasnet}, as the difficulty measure and dynamically chooses training samples within a mini-batch based on the loss threshold. By controlling the threshold, we can design a curriculum. 

\subsection{Training scheduler}

The key step of the training scheduler for our CL strategies apart from the self-paced one, is simple ––- we  divide the training data into subsets based on a difficulty measure, sort the subsets from “easy” to “hard”, and start training the TSE model parameters with the easiest subset. After a ﬁxed number of training epochs or the convergence of the training loss, the next subset is merged into the training portion. The model parameters are further updated using the expanded training portion. This process is repeated until all the subsets are merged into the training portion and until the model parameters are well-trained.

The training scheduler for the self-paced strategy is different from the ones for other proposed CL strategies. We form the training mini-batch using standard random sampling and compute the objective value for each sample within a mini-batch using the current model that is being trained. The curriculum is designed by adjusting the threshold that chooses the samples to be used for computing the averaged value of the objective function before calculating gradients. For the SNR-based objective function, like other CL strategies, we gradually decrease the threshold so that “easy” samples are dominantly used to compute the gradient initially, and more difficult data are also considered at the later stages of the training. This approach helps deal with difficult samples that are challenging to process at the early stage of learning, improving the stability of the training and thus the generalization ability of the network.

\section{Experiments and Results}

\label{sec:typestyle}
\subsection{Dataset}
\begin{table}[tb]
\centering
\caption{Dataset and its partitions for this study. No overlap existed between the test partition and the other two partitions at either the utterance or speaker level.}
\footnotesize
\label{tab:dataset_info}
\begin{tabular}{lcrr}
\toprule
\textbf{Dataset} & \textbf{Partition} & \textbf{\#utterances} & \textbf{\#speakers} \\
\hline
Libri2talker    & train              & 127,056                  & 1,172                  \\
                  & dev                & 2,344                    & 1,172                  \\
                  & test               & 6,000              & 40                      \\
\bottomrule
\end{tabular}
\vspace{-5mm}
\end{table}
To assess the proposed CL strategies for target speaker extraction, We employed the Libri2talker~\cite{xu2021target} configuration, a simulated 2-talker mixture audio dataset derived from the LibriSpeech dataset~\cite{panayotov2015librispeech}. Libri2talker expanded the Libri2mix~\cite{cosentino2020librimix} dataset by alternating the target and interfering speakers and by reusing each 2-talker mixture used twice. The statistics of each partition are shown in Table \ref{tab:dataset_info}.

\subsection{Experimental conditions}
We first brief the feature extraction. The sampling frequency used for this study was 16 KHz.
All reference speeches were either padded or segmented with a length of 15 seconds, while all mixture speeches clipped to 6 seconds. Speaker embeddings were extracted using an ECAPA-TDNN \cite{desplanques2020ecapa} pre-trained on the VoxCeleb2 dataset \cite{chung2018voxceleb2}. For computing the STFT of the mixture signal, we set the window length and hop size to 32 ms and 8 ms, respectively, with an FFT length of 512. 

Let us move on to the training hyper-parameter setup. The optimizer was Adam \cite{adam}, starting with a learning rate of 1e-3. Following the similar scheme from \cite{attention_is_not_what_i_need}, it gradually increased during the first 5000 batches, then applied a decay to reduce the learning rate, ensuring it did not fall below 1e-5. Each model was trained with different random seeds three times to obtain averaged results. In each mini-batch, the model processed 48 pairs of target and interfering speakers. 

The improvement in performance of the proposed CL strategies was evaluated using on an SDR calculation. The improvement in SDR (iSDR), measured as the relative increase in SDR compared to the mixture, was evaluated.
All the proposed CL strategies except the self-paced method have 2 phases. In the 1st phase, the TSE model was trained in 100 epochs on \textbf{“easy”} samples, which were chosen in accordance with the difficulty measurers described in Section 3. The 2nd phase built on the model of the 1st phase and continued training for 5 epochs. In the 2nd phase, all the samples of the training set instead of the selected ones were used. The details of the data selection for each CL are presented as follows: 
\begin{description}
\item[\textbf{Random}:] This method served as the baseline for the study. The TSE model was trained for the 100 epochs, where the training set was randomly sampled to create the mini-batches.
\item[\textbf{Gender}:] The “easy” samples were selected via different genders between the target and interfering speakers in the 1st phase, in accordance with the description in Section 3.1.1. The gender information of the speakers could be acquired via meta-labels provided within the dataset.
\item[\textbf{Speaker similarity}:] The “easy” samples were selected from the training partition with low speaker similarity between the target and interfering speakers via a threshold $\tau_{spk}$ at the 1st phase\footnote{Strictly speaking, there are 3 phases if training the speaker encoder is seen as 1 phase, but here it's treated as an external step.}. Consequently, three distinct values were assigned to  $\tau_{spk} \in [0.5, 0.6, 0.7]$, and they were fixed throughout the training process.
\item[\textbf{SDR}:] The “easy” samples for the 1st phase were selected based on the SDR threshold $\tau_{SDR}$ for each training sample. In this study, we systematically varied $\tau_{SDR}$ across a range from -5 dB to 5 dB with 1 dB increments, resulting in the evaluation of eleven conditions.
\item[\textbf{SNR}:] The “easy” samples for the 1st phase were selected based on the SNR threshold $\tau_{SNR}$ of each training sample\footnote{The TSE model used for the SNR calculation was from the baseline method. Similar to the treatment of the speaker encoder above, the SNR calculation was treated as an external step.}. Here, $\tau_{SNR}$ was varied from 0 dB to 10 dB, with increments of 5 dB, resulting in the evaluation of three conditions.
\item[\textbf{Self-paced}:] 
The self-paced CL requires a more sophisticated training scheduler comprising five distinct phases across the 100 epochs. We only back-propagate the gradients from the data in a batch where the loss (SNR, specifically) was greater than a threshold $\tau_{SP}$. Initially, the model underwent 1 epoch of training with the entirety of the training data. Subsequently, the training proceeded through 3 phases demarcated by triples, denoting the (\emph{start\_epoch}, \emph{end\_epoch}, $\tau_{SP}$) configurations. Specifically, these phases were instantiated as (1, 30, 10 dB), (31, 60, 5 dB), and (61, 80, 0 dB), respectively. Finally, the model underwent the last phase comprising 20 epochs, utilizing the entire training data.
\end{description}
\subsection{Results}
\label{sec:1}
\noindent 
Results on the test set using random selection, Gender, Speaker similarity, SDR, and SNR based CL are shown in Table \ref{tab:combined_results}. Note that the 1st phase took 100 epochs and the 2nd phase was conducted for a further 5 epochs. The optimal thresholds $\tau$ for each difficulty measure is tuned on the dev set. Below are the findings.

\noindent\textbf{Random:} 
Training for 100 epochs with random sample selection reached a 12.50 dB SDR, with an additional 5 epochs yielding a minor 0.07 dB increase. This served as the baseline.

\vspace{1mm}
\noindent\textbf{Gender:} For convenience, in the table we use $\tau_{gender} \in [\text{Same, Different}]$ to indicate if the gender between the target and interfering speakers are same. The gender-based CL method resulted in worse than with random selection, with the resulting iSDR returned by the 2nd phase being -1.1 dB compared with the baseline.

\vspace{1mm}
\noindent\textbf{Speaker similarity:} 
 The results underscore the efficacy of cosine distance-based speaker similarity as a reliable difficulty measure within the proposed CL framework. Remarkably, setting $\tau_{spk}$ to 0.6 resulted in a notable improvement of 13.44 dB.

\begin{figure}[ht]
    \centering
        \centering
        \includegraphics[width=0.85\columnwidth]{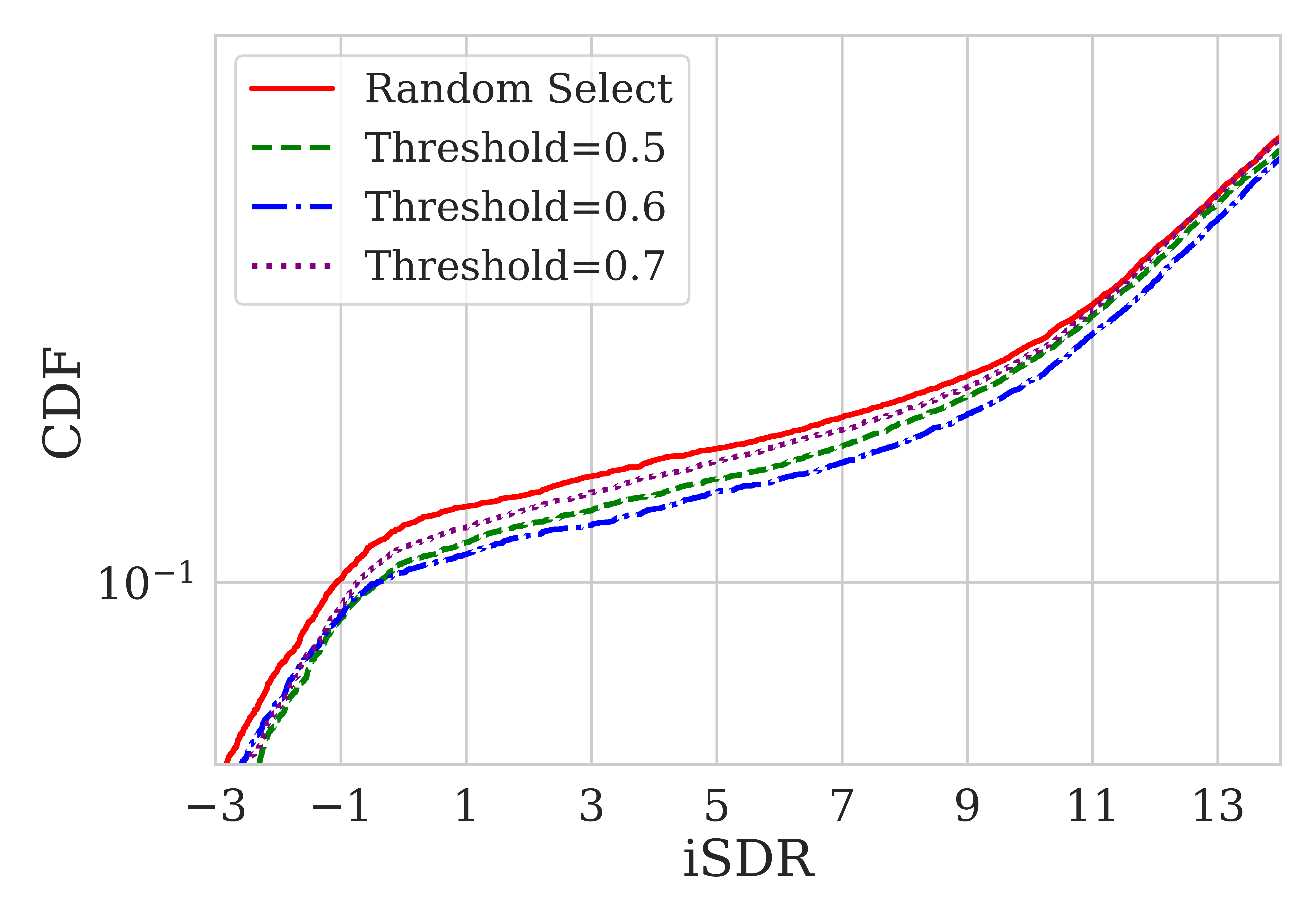}
        \caption{Cumulative density function of prediction results from the TSE model based on speaker similarity. Here, larger CDF values indicates higher selected proportion of corresponding iSDR(dB) in the corresponding system. ``Threshold" corresponds to $\tau_{spk}$ in Section 4.2. ``Random Select" is the baseline.}
        \label{fig:kde}
    \label{fig:test}
    \vspace{-2mm}
\end{figure}

We computed iSDR for all samples in the test set to gain further insights on how the CL method using the speaker similarity difficulty measurer improved the TSE model in the 2nd phase compared with the baseline. We then performed kernel density estimation on the obtained SDRs and derived cumulative density functions (CDF), as depicted in Figure \ref{fig:kde}. 
The 'Random Select' method demonstrates a higher CDF probability than the CL methods when the iSDR is low, indicating that it contains more samples with minor or negative improvement compared to the proposed method. As the iSDR increases, the gap between them becomes smaller, suggesting that the proposed CL methods includes more samples with large improvement.

\begin{table}[htbp]
\footnotesize
\centering
\caption{iSDR(dB) results on the test set using the CL-based TSE. ``Used data" shows the percentage of the used data among the whole dataset in the 1st phase. The 1st phase results were from the systems trained using “easy” samples. The 2nd phase results were from the systems additionally trained using all samples based on the model weights in the 1st phase.  }
\label{tab:combined_results}
\begin{tabular}{lcrrr}
\toprule
& \textbf{Optimal $\tau$ }& \textbf{Used data} & \textbf{1st phase} & \textbf{2nd phase}  \\ \hline
{$random$} & - & 100\% & 12.50 & 12.57 \\ 
 \hline
{$\tau_{gender}$} & Different & 50.0\% & 7.64 & 11.40 \\ 
{$\tau_{spk}$} & 0.6 & 83.8\% & 13.03 & \textbf{13.44} \\ 
{$\tau_{SDR}$} & 1 & 44.3\% & 8.75 & 13.40 \\
{$\tau_{SNR}$} & 10 & 82.4\% & 12.79 & 12.99 \\ 

 \bottomrule
\end{tabular}
\vspace{-2mm}
\end{table}

\begin{table}[t]
\centering
\footnotesize
\caption{iSDR(dB) results on the test set using the self-paced CL divided by 5 phases, as described in Section 4.2.}
\label{tab:self_th}
\begin{tabular}{crrrr}
\toprule
\textbf{1st phase} & \textbf{2nd phase} & \textbf{3rd phase}  & \textbf{4th phase} & \textbf{5th phase} \\
\hline
3.30  & 7.79  & 10.93   & 12.66  & \textbf{13.54}  \\
\bottomrule
\end{tabular}
\end{table}

\begin{figure}[h]
\vspace{-3mm}
    \centering
        \centering
        \includegraphics[width=0.85\columnwidth]{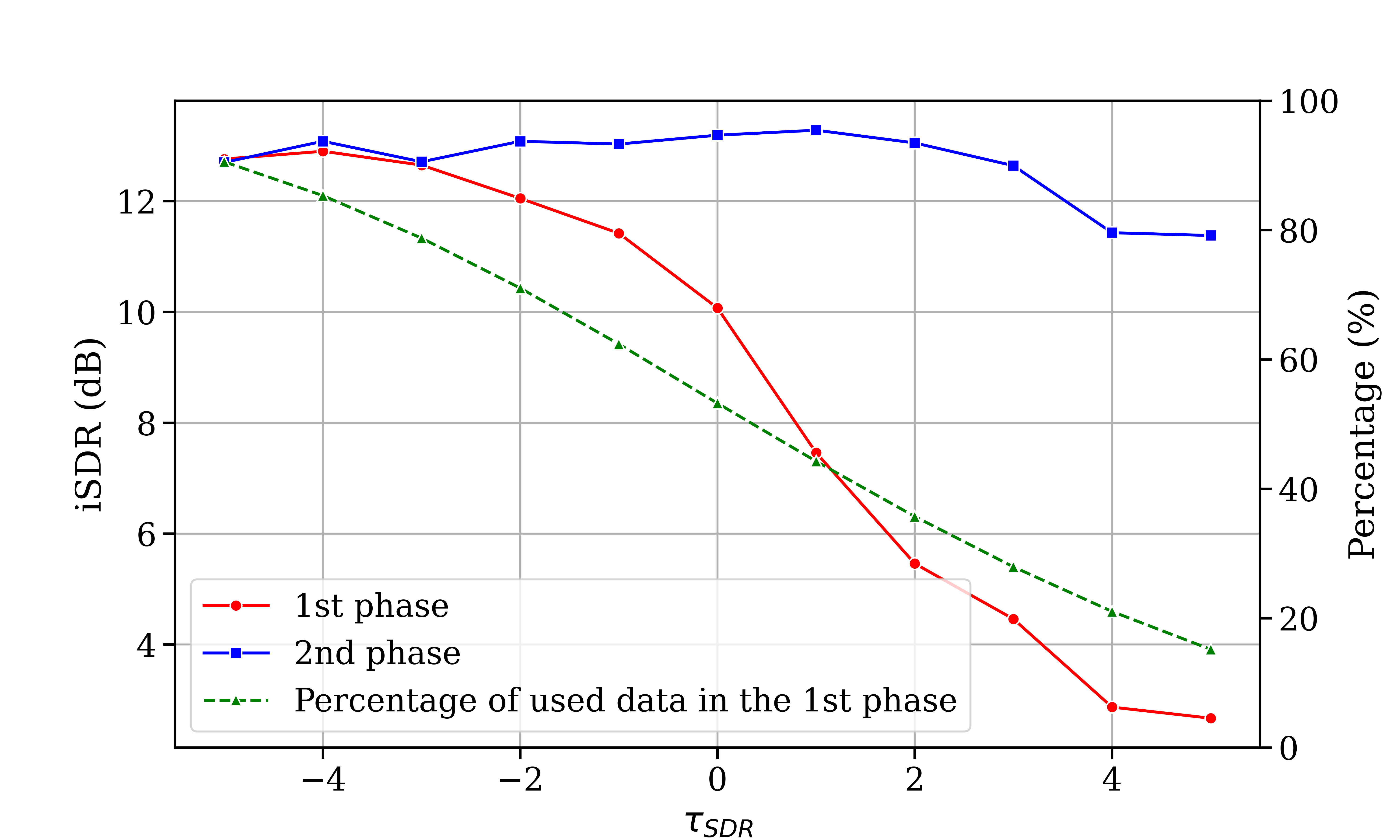}
        \caption{iSDR changes on the dev set with $\tau_{SDR}$  in two training phases, along with data usage percentage in the 1st phase. }
        \label{fig:sdrcl}
    \vspace{-2mm}
\end{figure}

\vspace{1mm}
\noindent 
\textbf{SDR}: 
When $\tau_{SDR}$  is set to 1, the iSDR increases from 8.75 in the 1st phase to 13.40 in the 2nd phase. To further validate this trend, we plot the iSDR changes on dev set across different thresholds for the 1st and 2nd phases in Figure \ref{fig:sdrcl}, along with the percentage of data used in the 1st phase. The first observation indicates a consistent increase in iSDR across all thresholds, confirming that gradually exposing the model to more challenging samples is effective for achieving better performance. Additionally, the selection of the optimal threshold to balance between 'easier' and more complex data is sensitive and crucial.

\vspace{1mm}
\noindent 
\textbf{SNR}: While the results using speaker similarity as the difficulty measure achieves promising results, utilizing similar data proportion scheme, the SNR-based difficulty measure achieves rather marginal iSDR improvement. This further enhances the importance of designing proper difficulty measures for the CL.
\noindent \textbf{Self-paced}: The results using the self-paced CL are shown in Table \ref{tab:self_th}. After the five designed phases specific for this strategy, 1.04 dB relative improvement over the baseline is observed, which is the best across all the proposed methods.
In the 4th phase, the model only back-propagates the gradients when the SNR loss is no less than 0. At this point, without considering training with the complex data, improvements over the baseline have already been achieved. The negative SNR loss may correspond to the extraction of the wrong speaker, and setting such cases to later phases is beneficial.

In summary, the self-paced CL approach stands out with the best iSDR improvement. Meanwhile, it is worth noting that compared with the alternatives, this method may entail greater complexity in design and training. If one wants to maintain a balance between performance and engineering efforts, the speaker similarity-based CL can be a pragmatic choice. Notably, it yielded the second-best iSDR across all the methods and is relatively straightforward to utilize.

\begin{table}[t]

\centering

\caption{iSDR(dB) of CL methods with different architecture.}
\footnotesize
\label{tab:netwroks}
\begin{tabular}{crrrr}
\toprule
\textbf{Networks} & \textbf{random} & \textbf{similarity(0.6)} & \textbf{self-paced} \\
\hline
Naive-BLSTM & 8.09  & 8.30  & \textbf{9.60}  \\
SpeakerBeam\cite{SpeakerBeam2019} & 9.47   & 9.78  & \textbf{9.88}  \\
VoiceFilter \cite{wang2018voicefilter} & 6.90 & 7.17  & \textbf{7.54}  \\
\bottomrule
\end{tabular}
\vspace{-4mm}
\end{table}

Table \ref{tab:netwroks} presents the results obtained using two CL strategies with different architectures using the best parameters above. Naive-BLSTM, utilizing the same input and output manner as the Conformer TSE, merely replaces the 4 Conformer blocks with a 2-layer 512 BLSTM. This model with self-paced method significantly improved 1.51dB than the random selection method. SpeakerBeam employs the same structure as \cite{SpeakerBeam2019}, while VoiceFilter simply substitutes d-vector in \cite{wang2018voicefilter} with the speaker embedding based on ECAPA-TDNN described above. Reflected in the table, the two selected CL methods consistently shows improvement compared to random selection.

\section{Conclusion}
\label{sec:conc}
This paper has presented several CL strategies tailored for modern neural TSE models based on multiple well-established thresholds for difficulty measures. All strategies share a common underlying principle of gradually introducing more challenging training data, each employs a difficulty measure based on the thresholds to discern between "easy" and difficult samples. 
Through extensive evaluation on the Libri2talker, our study has showed that our proposed CL strategies are able to substantially enhance the performance of de facto TSE models. 

\vfill\pagebreak

\section{Acknowledgements}
This study is partially supported by MEXT KAKENHI Grants (21H04906) and JST, the establishment of university fellowships towards the creation of science technology innovation (JPMJFS2136).

\bibliographystyle{IEEEtran}
\bibliography{main}

\begin{thebibliography}{10}
\providecommand{\url}[1]{#1}
\csname url@samestyle\endcsname
\providecommand{\newblock}{\relax}
\providecommand{\bibinfo}[2]{#2}
\providecommand{\BIBentrySTDinterwordspacing}{\spaceskip=0pt\relax}
\providecommand{\BIBentryALTinterwordstretchfactor}{4}
\providecommand{\BIBentryALTinterwordspacing}{\spaceskip=\fontdimen2\font plus
\BIBentryALTinterwordstretchfactor\fontdimen3\font minus
  \fontdimen4\font\relax}
\providecommand{\BIBforeignlanguage}[2]{{%
\expandafter\ifx\csname l@#1\endcsname\relax
\typeout{** WARNING: IEEEtran.bst: No hyphenation pattern has been}%
\typeout{** loaded for the language `#1'. Using the pattern for}%
\typeout{** the default language instead.}%
\else
\language=\csname l@#1\endcsname
\fi
#2}}
\providecommand{\BIBdecl}{\relax}
\BIBdecl

\bibitem{Zmolikova_Spkbeam_STSP19}
K.~Žmolíková, M.~Delcroix, K.~Kinoshita, T.~Ochiai, T.~Nakatani, L.~Burget,
  and J.~Černocký, ``Speakerbeam: Speaker aware neural network for target
  speaker extraction in speech mixtures,'' \emph{IEEE Journal of Selected
  Topics in Signal Processing}, vol.~13, no.~4, pp. 800--814, 2019.

\bibitem{Wang2019}
\BIBentryALTinterwordspacing
Q.~Wang, H.~Muckenhirn, K.~Wilson, P.~Sridhar, Z.~Wu, J.~R. Hershey, R.~A.
  Saurous, R.~J. Weiss, Y.~Jia, and I.~L. Moreno, ``{VoiceFilter: Targeted
  Voice Separation by Speaker-Conditioned Spectrogram Masking},'' in
  \emph{Proc. Interspeech 2019}, 2019, pp. 2728--2732. [Online]. Available:
  \url{http://dx.doi.org/10.21437/Interspeech.2019-1101}
\BIBentrySTDinterwordspacing

\bibitem{fu2019metricgan}
S.-W. Fu, C.-F. Liao, Y.~Tsao, and S.-D. Lin, ``Metricgan: Generative
  adversarial networks based black-box metric scores optimization for speech
  enhancement,'' in \emph{International Conference on Machine Learning}.\hskip
  1em plus 0.5em minus 0.4em\relax PMLR, 2019, pp. 2031--2041.

\bibitem{zhao2022target}
Z.~Zhao, D.~Yang, R.~Gu, H.~Zhang, and Y.~Zou, ``{Target Confusion in
  End-to-end Speaker Extraction: Analysis and Approaches},'' in \emph{Proc.
  Interspeech 2022}, 2022, pp. 5333--5337.

\bibitem{wang2021survey}
X.~Wang, Y.~Chen, and W.~Zhu, ``A survey on curriculum learning,'' \emph{IEEE
  Transactions on Pattern Analysis and Machine Intelligence}, vol.~44, no.~9,
  pp. 4555--4576, 2021.

\bibitem{soviany2022curriculum}
P.~Soviany, R.~T. Ionescu, P.~Rota, and N.~Sebe, ``Curriculum learning: A
  survey,'' \emph{International Journal of Computer Vision}, vol. 130, no.~6,
  pp. 1526--1565, 2022.

\bibitem{Guo_2018_ECCV}
S.~Guo, W.~Huang, H.~Zhang, C.~Zhuang, D.~Dong, M.~R. Scott, and D.~Huang,
  ``Curriculumnet: Weakly supervised learning from large-scale web images,'' in
  \emph{Proceedings of the European Conference on Computer Vision (ECCV)},
  September 2018.

\bibitem{jiang2014easy}
L.~Jiang, D.~Meng, T.~Mitamura, and A.~G. Hauptmann, ``Easy samples first:
  Self-paced reranking for zero-example multimedia search,'' in
  \emph{Proceedings of the 22nd ACM international conference on Multimedia},
  2014, pp. 547--556.

\bibitem{platanios-etal-2019-competence}
\BIBentryALTinterwordspacing
E.~A. Platanios, O.~Stretcu, G.~Neubig, B.~Poczos, and T.~Mitchell,
  ``Competence-based curriculum learning for neural machine translation,'' in
  \emph{Proceedings of the 2019 Conference of the North {A}merican Chapter of
  the Association for Computational Linguistics: Human Language Technologies,
  Volume 1 (Long and Short Papers)}.\hskip 1em plus 0.5em minus 0.4em\relax
  Minneapolis, Minnesota: Association for Computational Linguistics, Jun. 2019,
  pp. 1162--1172. [Online]. Available: \url{https://aclanthology.org/N19-1119}
\BIBentrySTDinterwordspacing

\bibitem{tay-etal-2019-simple}
\BIBentryALTinterwordspacing
Y.~Tay, S.~Wang, A.~T. Luu, J.~Fu, M.~C. Phan, X.~Yuan, J.~Rao, S.~C. Hui, and
  A.~Zhang, ``Simple and effective curriculum pointer-generator networks for
  reading comprehension over long narratives,'' in \emph{ACL}, A.~Korhonen,
  D.~Traum, and L.~M{\`a}rquez, Eds.\hskip 1em plus 0.5em minus 0.4em\relax
  Florence, Italy: ACL, Jul. 2019, pp. 4922--4931. [Online]. Available:
  \url{https://aclanthology.org/P19-1486}
\BIBentrySTDinterwordspacing

\bibitem{Progressivenina}
Z.~Nian, Y.-H. Tu, J.~Du, and C.-H. Lee, ``A progressive learning approach to
  adaptive noise and speech estimation for speech enhancement and noisy speech
  recognition,'' in \emph{ICASSP 2021 - 2021 IEEE International Conference on
  Acoustics, Speech and Signal Processing (ICASSP)}, 2021, pp. 6913--6917.

\bibitem{tan2021efficientnetv2}
M.~Tan and Q.~Le, ``Efficientnetv2: Smaller models and faster training,'' in
  \emph{International conference on machine learning}.\hskip 1em plus 0.5em
  minus 0.4em\relax PMLR, 2021, pp. 10\,096--10\,106.

\bibitem{li2020speech}
A.~Li, M.~Yuan, C.~Zheng, and X.~Li, ``Speech enhancement using progressive
  learning-based convolutional recurrent neural network,'' \emph{Applied
  Acoustics}, vol. 166, p. 107347, 2020.

\bibitem{gao16_interspeech}
T.~Gao, J.~Du, L.-R. Dai, and C.-H. Lee, ``{SNR-Based Progressive Learning of
  Deep Neural Network for Speech Enhancement},'' in \emph{Proc. Interspeech
  2016}, 2016, pp. 3713--3717.

\bibitem{Ranjan2018ASRCL}
S.~Ranjan and J.~H.~L. Hansen, ``Curriculum learning based approaches for noise
  robust speaker recognition,'' \emph{IEEE/ACM Transactions on Audio, Speech,
  and Language Processing}, vol.~26, no.~1, pp. 197--210, 2018.

\bibitem{sinha2022speaker}
R.~Sinha, M.~Tammen, C.~Rollwage, and S.~Doclo, ``Speaker-conditioning
  single-channel target speaker extraction using conformer-based
  architectures,'' in \emph{2022 International Workshop on Acoustic Signal
  Enhancement (IWAENC)}.\hskip 1em plus 0.5em minus 0.4em\relax IEEE, 2022, pp.
  1--5.

\bibitem{cirm}
D.~S. Williamson, Y.~Wang, and D.~Wang, ``Complex ratio masking for monaural
  speech separation,'' \emph{IEEE/ACM Transactions on Audio, Speech, and
  Language Processing}, vol.~24, no.~3, pp. 483--492, 2016.

\bibitem{hu20g_interspeech}
Y.~Hu, Y.~Liu, S.~Lv, M.~Xing, S.~Zhang, Y.~Fu, J.~Wu, B.~Zhang, and L.~Xie,
  ``{DCCRN: Deep Complex Convolution Recurrent Network for Phase-Aware Speech
  Enhancement},'' in \emph{Proc. Interspeech 2020}, 2020, pp. 2472--2476.

\bibitem{gulati2020conformer}
A.~Gulati, J.~Qin, C.-C. Chiu, N.~Parmar, Y.~Zhang, J.~Yu, W.~Han, S.~Wang,
  Z.~Zhang, Y.~Wu \emph{et~al.}, ``Conformer: Convolution-augmented transformer
  for speech recognition,'' \emph{Proc. Interspeech}, pp. 5036--5040, 2020.

\bibitem{desplanques2020ecapa}
B.~Desplanques, J.~Thienpondt, and K.~Demuynck, ``Ecapa-tdnn: Emphasized
  channel attention, propagation and aggregation in tdnn based speaker
  verification,'' \emph{Proc. Interspeech 2020}, pp. 3830--3834, 2020.

\bibitem{convtasnet}
Y.~Luo and N.~Mesgarani, ``Conv-tasnet: Surpassing ideal time–frequency
  magnitude masking for speech separation,'' \emph{IEEE/ACM Transactions on
  Audio, Speech, and Language Processing}, vol.~27, no.~8, pp. 1256--1266,
  2019.

\bibitem{xu2021target}
C.~Xu, W.~Rao, J.~Wu, and H.~Li, ``Target speaker verification with selective
  auditory attention for single and multi-talker speech,'' \emph{IEEE/ACM
  Transactions on audio, speech, and language processing}, vol.~29, pp.
  2696--2709, 2021.

\bibitem{panayotov2015librispeech}
V.~Panayotov, G.~Chen, D.~Povey, and S.~Khudanpur, ``Librispeech: an asr corpus
  based on public domain audio books,'' in \emph{2015 IEEE international
  conference on acoustics, speech and signal processing (ICASSP)}.\hskip 1em
  plus 0.5em minus 0.4em\relax IEEE, 2015, pp. 5206--5210.

\bibitem{cosentino2020librimix}
J.~Cosentino, M.~Pariente, S.~Cornell, A.~Deleforge, and E.~Vincent,
  ``Librimix: An open-source dataset for generalizable speech separation,''
  2020.

\bibitem{chung2018voxceleb2}
J.~S. Chung, A.~Nagrani, and A.~Zisserman, ``Voxceleb2: Deep speaker
  recognition,'' in \emph{INTERSPEECH}, 2018.

\bibitem{adam}
D.~P. Kingma and J.~Ba, ``Adam: {A} method for stochastic optimization,'' in
  \emph{3rd International Conference on Learning Representations, {ICLR} 2015,
  San Diego, CA, USA, May 7-9, 2015, Conference Track Proceedings}, 2015.

\bibitem{attention_is_not_what_i_need}
\BIBentryALTinterwordspacing
A.~Vaswani, N.~Shazeer, N.~Parmar, J.~Uszkoreit, L.~Jones, A.~N. Gomez, L.~u.
  Kaiser, and I.~Polosukhin, ``Attention is all you need,'' in \emph{Advances
  in Neural Information Processing Systems}, I.~Guyon, U.~V. Luxburg,
  S.~Bengio, H.~Wallach, R.~Fergus, S.~Vishwanathan, and R.~Garnett, Eds.,
  vol.~30.\hskip 1em plus 0.5em minus 0.4em\relax Curran Associates, Inc.,
  2017. [Online]. Available:
  \url{https://proceedings.neurips.cc/paper/2017/file/3f5ee243547dee91fbd053c1c4a845aa-Paper.pdf}
\BIBentrySTDinterwordspacing

\bibitem{SpeakerBeam2019}
K.~Žmolíková, M.~Delcroix, K.~Kinoshita, T.~Ochiai, T.~Nakatani, L.~Burget,
  and J.~Černocký, ``Speakerbeam: Speaker aware neural network for target
  speaker extraction in speech mixtures,'' \emph{IEEE Journal of Selected
  Topics in Signal Processing}, vol.~13, no.~4, pp. 800--814, 2019.

\bibitem{wang2018voicefilter}
Q.~Wang, H.~Muckenhirn, K.~Wilson, P.~Sridhar, Z.~Wu, J.~R. Hershey, R.~A.
  Saurous, R.~J. Weiss, Y.~Jia, and I.~L. Moreno, ``{VoiceFilter: Targeted
  Voice Separation by Speaker-Conditioned Spectrogram Masking},'' in
  \emph{Proc. Interspeech}, 2019, pp. 2728--2732.

\end{thebibliography}

\end{document}